\documentclass[aip, amsmath,amssymb, reprint]{revtex4-1}%twocolumn,   apl,

\usepackage{multirow}
\usepackage{bigstrut}
\usepackage{amssymb}
\usepackage{amsmath}
\usepackage{graphicx}
\usepackage{dcolumn}
\usepackage{bm}
\usepackage{CJK}
\usepackage{relsize}
\usepackage{cancel}
\usepackage{color}
\usepackage{subfig}

\begin{document}

\title{Statistical exchange-coupling errors and the practicality of scalable silicon donor qubits}

\author{Yang Song}\email{ysong128@umd.edu}\affiliation{Condensed Matter Theory Center, Department of Physics, University of Maryland, College Park, MD 20742, USA}\affiliation{Joint Quantum Institute, University of Maryland, College Park, MD 20742, USA}
\author{S.~Das Sarma}
\affiliation{Condensed Matter Theory Center, Department of Physics, University of Maryland, College Park, MD 20742, USA}\affiliation{Joint Quantum Institute, University of Maryland, College Park, MD 20742, USA}

\begin{abstract}
Recent experimental efforts have led to considerable interest in donor-based localized electron spins in Si as viable qubits for a scalable silicon quantum computer. With the use of isotopically purified $^{28}$Si and the realization of extremely long spin coherence time in single-donor electrons, the recent experimental focus is on two-coupled donors with the eventual goal of a scaled-up quantum circuit. Motivated by this development, we simulate the statistical distribution of the exchange coupling $J$ between a pair of donors under realistic donor placement straggles, and quantify the errors relative to the intended $J$ value. With $J$ values in a broad range of donor-pair separation ($5<|\mathbf{R}|<60$ nm), we work out various cases systematically, for a target donor separation $\mathbf{R}_0$ along the [001], [110] and [111] Si crystallographic directions, with $|\mathbf{R}_0|=10, 20$ or 30 nm and standard deviation $\sigma_R=1, 2, 5$ or 10 nm. Our extensive theoretical results demonstrate the great challenge for a prescribed $J$ gate even with just a donor pair, a first step for any scalable Si-donor-based quantum computer.
\end{abstract}
\maketitle

Solid state implementation of the quantum circuits based on silicon (Si) donor spins was proposed drawing on its attractive scale-up potential and long spin coherence time \cite{Kane_Nat98}. After nearly two decades of continued experimental improvements as well as theoretical investigations, impressive donor-based single-qubit performance has been achieved with very long coherence time \cite{Morello_Nature10, Tyryshkin_NatMat12, Pla_Nature12, Pla_Nature13, Saeedi_Science13, Zwanenburg_RMP13, Muhonen_NatNano14, Freer_arxiv16}.  Single qubit gates have been well established in multiple cases. However, as remarkable as these feats are, nonexistent as yet is a core necessary ingredient for reaching large scale circuits, namely, the inter-qubit coupling. Both one- and two-qubit gates are essential for quantum computing, and the current work is on two-qubit gates in donor-based electron spin qubits in Si. Fortunately, the iterative nature of circuit programming enables us to focus just on the two-qubit coupling, only after which the intrinsic scaling advantage of the solid state platform can be fully realized \cite{Kane_Nat98, Nielsen_Chuang_book}. The dominant coupling between two donors is the electron exchange coupling where experimental effort has started \cite{Dehollain_PRL14, Gonzales_NanoLett14, Weber_NatNano14}, and we focus on this mechanism.

One important feature for exchange coupling ($J$) between Si donor electrons is its oscillation with donor separation ($\mathbf{R}$) changing over the order of a lattice constant, due to the multi-valley Si conduction band, which was first pointed out in Refs.~\onlinecite{Koiller_PRL01, Koiller_PRB02} in the context of Si quantum computing. This feature enhances the sensitivity of $J$ on $R$ from the exponential decay length $r_B\sim 2$ nm (Bohr radius in Si) to $a\sim 0.5$ nm (lattice constant in Si).   Another inevitable feature in donor-based qubits is the randomness or uncertainty in donor placement, or, the ``straggling'' as it is called in the ion implantation literature \cite{Zwanenburg_RMP13}. These two features combined lead to an inherent randomness in the resulting $J$ values, which in turn leads to exchange gate errors via $\int J(t) dt$.  Currently the single-ion implantation technique has reduced straggle from about 10 nm \cite{Jamieson_APL05} to 5.8 nm \cite{Jacob_PRL16} with ion energy lowered from 14 to 6 keV, with a hope to reach 0.5 nm by ultralow cooling. Scanning tunneling microscopy (STM) combined with hydrogen lithography demonstrates $\pm 1$  lattice site accuracy in atomic placements\cite{Fuechsle_NatNanotec12, Weber_Science12}, but the yield is very low and the implantation process very slow.  We mention that even  single qubit control in STM-fabricated donor systems has not yet been experimentally demonstrated.

The impact of $J$ uncertainty \cite{Koiller_PRL01, Koiller_PRB02} on the practicality of multi-donor qubit gates has been recently investigated \cite{Gamble_PRB15, Mohiyaddin_PRB16}. These works aimed to find regions where $J$ is bounded from below, with a large probability ($>0.9$) to far exceed the thermal noise \cite{Gamble_PRB15} as well as the hyperfine coupling and Zeeman splitting energies in odd-number donor chains \cite{Mohiyaddin_PRB16}. While this probability goal is not sufficiently high for the stringent requirements in quantum computation, it sets an acceptable starting point given the expected steady future improvements. However, the serious lack of attention so far is on the estimation of $J$-gate error itself, not only its lower bound, due to straggling. A workable quantum circuit necessitates the capacity to design the two-qubit coupling as close as its intended target value, which becomes even more critical for a grid of qubit units in need of synchronization. This ability is also required in order to prescribe a single-triplet qubit gate via exchange coupling \cite{Levy_PRL02}.

In this paper, we set out to systematically quantify the statistical distribution of $J$ values for a series of realistic donor separations, $\mathbf{R}_0$. We highlight the non-negligible likelihood of $J$ falling into a large neighborhood of the intended $J_0\equiv J(\mathbf{R}_0)$, across a range of realistic straggle deviations ($\sigma_R$) arising from ion implantation. In order to accomplish this, we continuously map out calculated $J(\mathbf{R})$ over the broad parameter range of $5< |\mathbf{R}|< 60$ nm. Then an arbitrary straggle distribution can be modeled by superimposing it onto the 3D function $J(\mathbf{R})$.

\begin{figure}[!htbp]
\centering
\includegraphics[width=8.5cm]{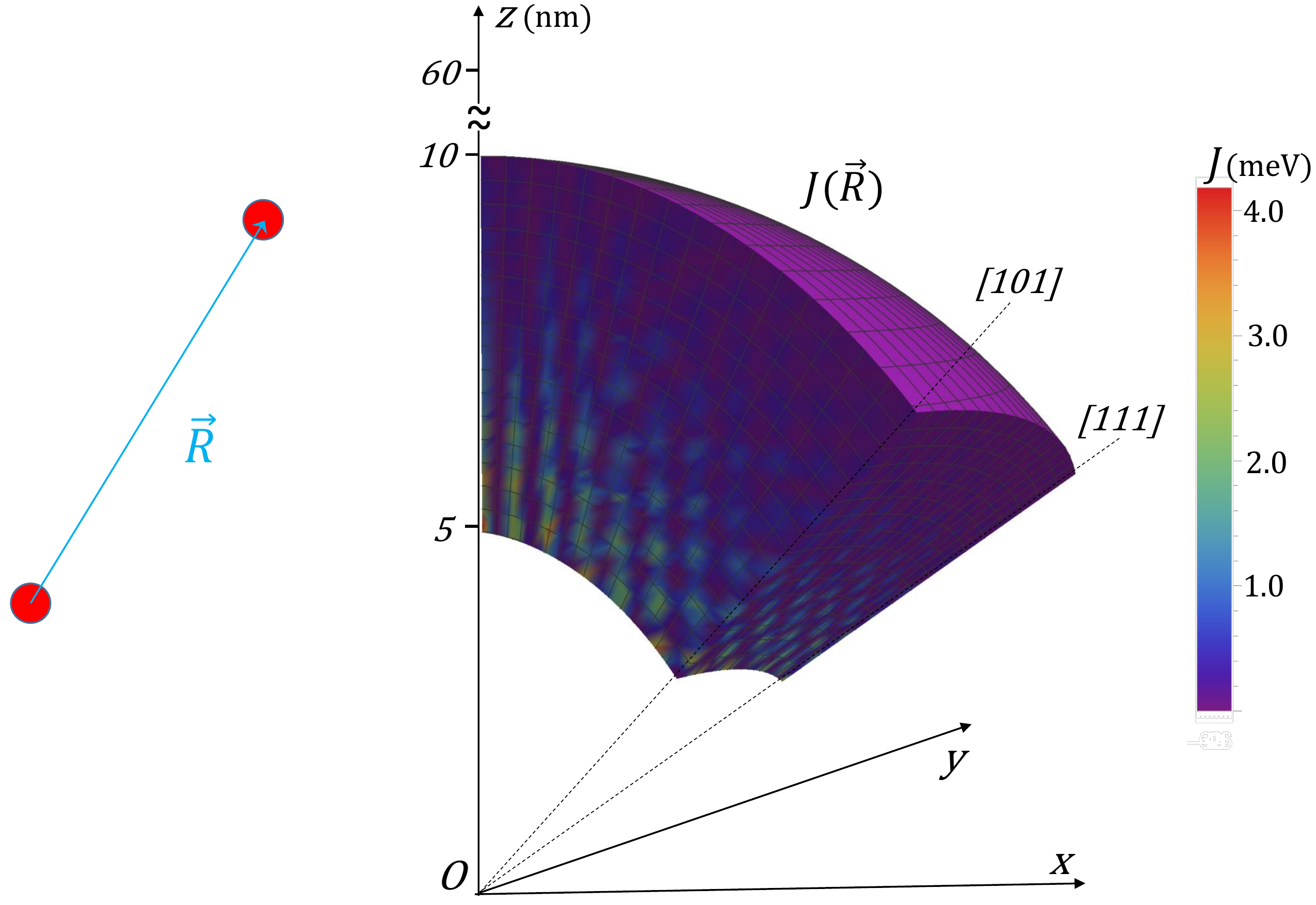}
\caption{ Exchange coupling $J(\mathbf{R})$ as a function of two-donor separation $\mathbf{R}$, in an irreducible 1/48 wedge. We have $J(\mathbf{R})$ mapped over all the $5<|\mathbf{R}|<60$ nm region. To visually discern the variation of $J$ on linear scale, we only show here a snapshot of the surface of the 3D wedge between $|\mathbf{R}|=5$ and 10 nm surfaces.
}\label{fig:fig_1}
\end{figure}

The exchange interaction between two electrons of two shallow donors in Si is calculated with the effective mass approach, taking into account  six anisotropic ellipsoidal valleys of the conduction band in Si crystal \cite{Kohn_SSP57}. All interaction Hamiltonian terms between electrons, the electron and ion, and ions are taken into account \cite{Herring_book66}, which ensures the correct sign for triplet-singlet energy ordering. The six-valley states enable us to fold any $\mathbf{R}$ between two donors into a 1/48 wedge (see Fig.~\ref{fig:fig_1}) for efficient irreducible computation. Both the Heitler-London approximation and the bilinear Heisenberg spin Hamiltonian ($J \mathbf{s}_1\cdot \mathbf{s}_2$) are valid only for $R$ not small compared to $r_B$, so we take a lower bound $R>5$ nm. The upper bound is set to be 60 nm (note for visibility resolution, only $5<R<10$ nm is shown in Fig.~\ref{fig:fig_1}).

\begin{figure}[!htbp]
\centering
  \subfloat
  {\includegraphics[width=8.5cm]{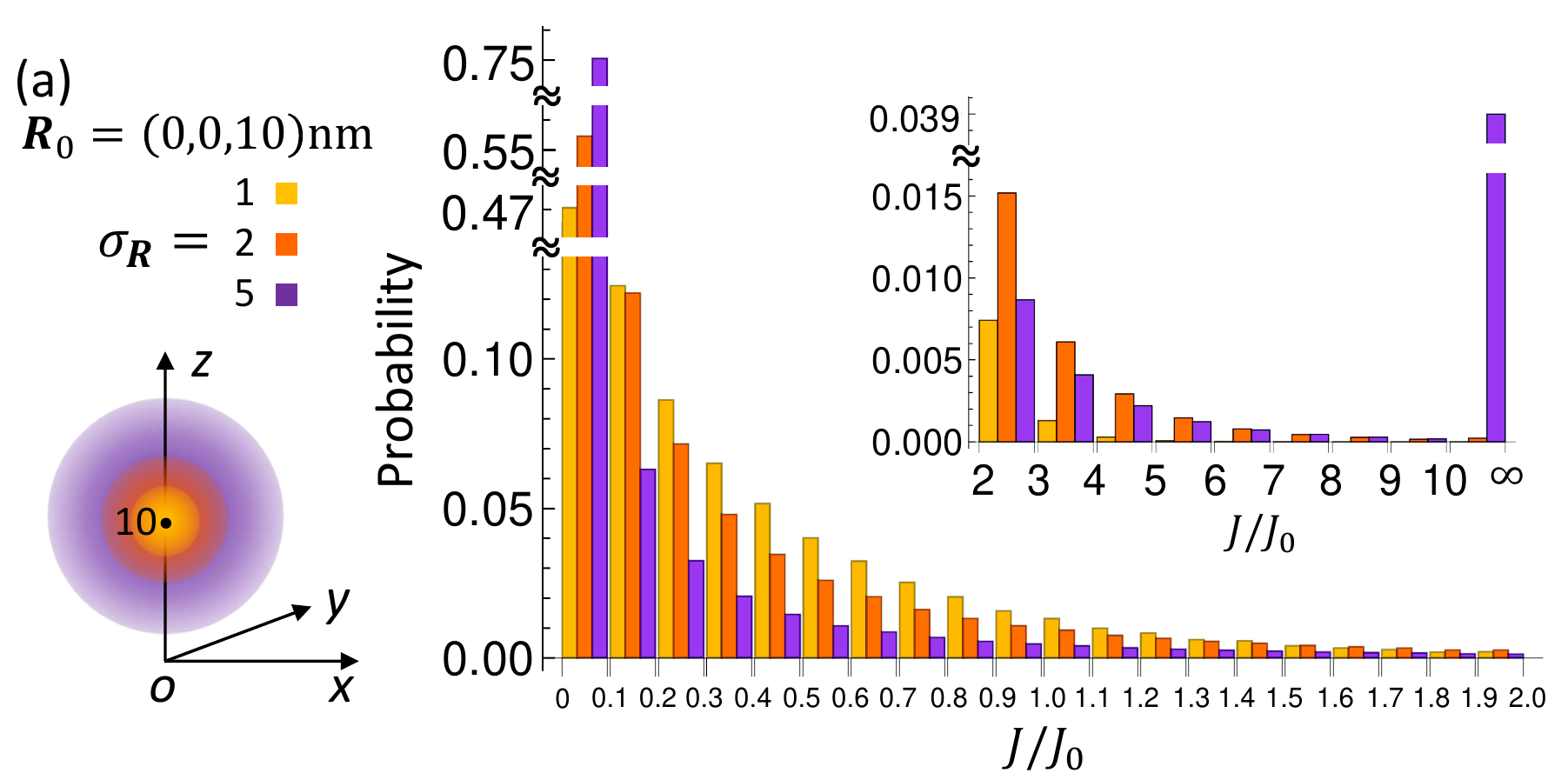}}
  \\
  \subfloat
  {\includegraphics[width=8.5cm]{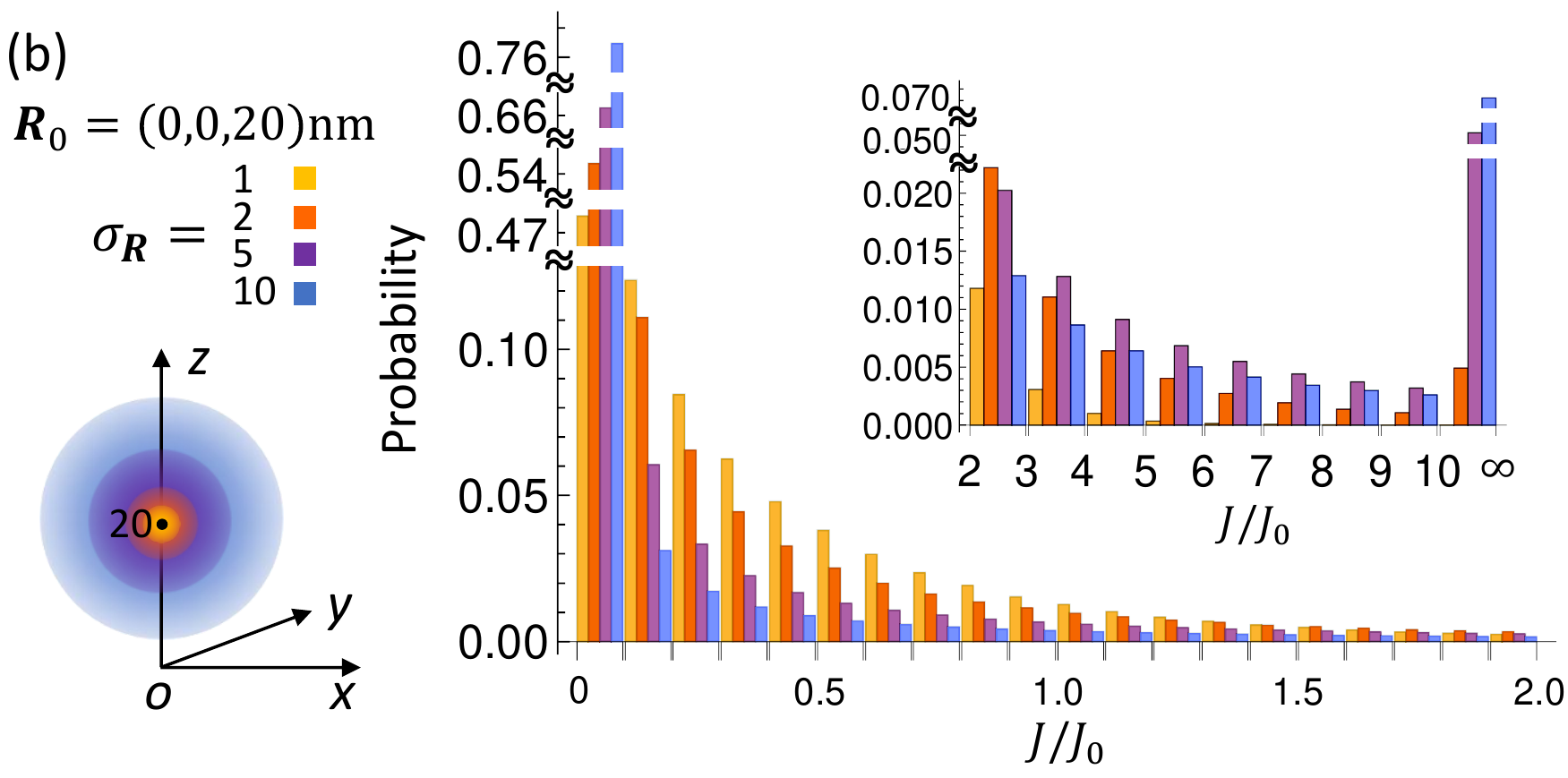}}
  \\
   \subfloat
  {\includegraphics[width=8.5cm]{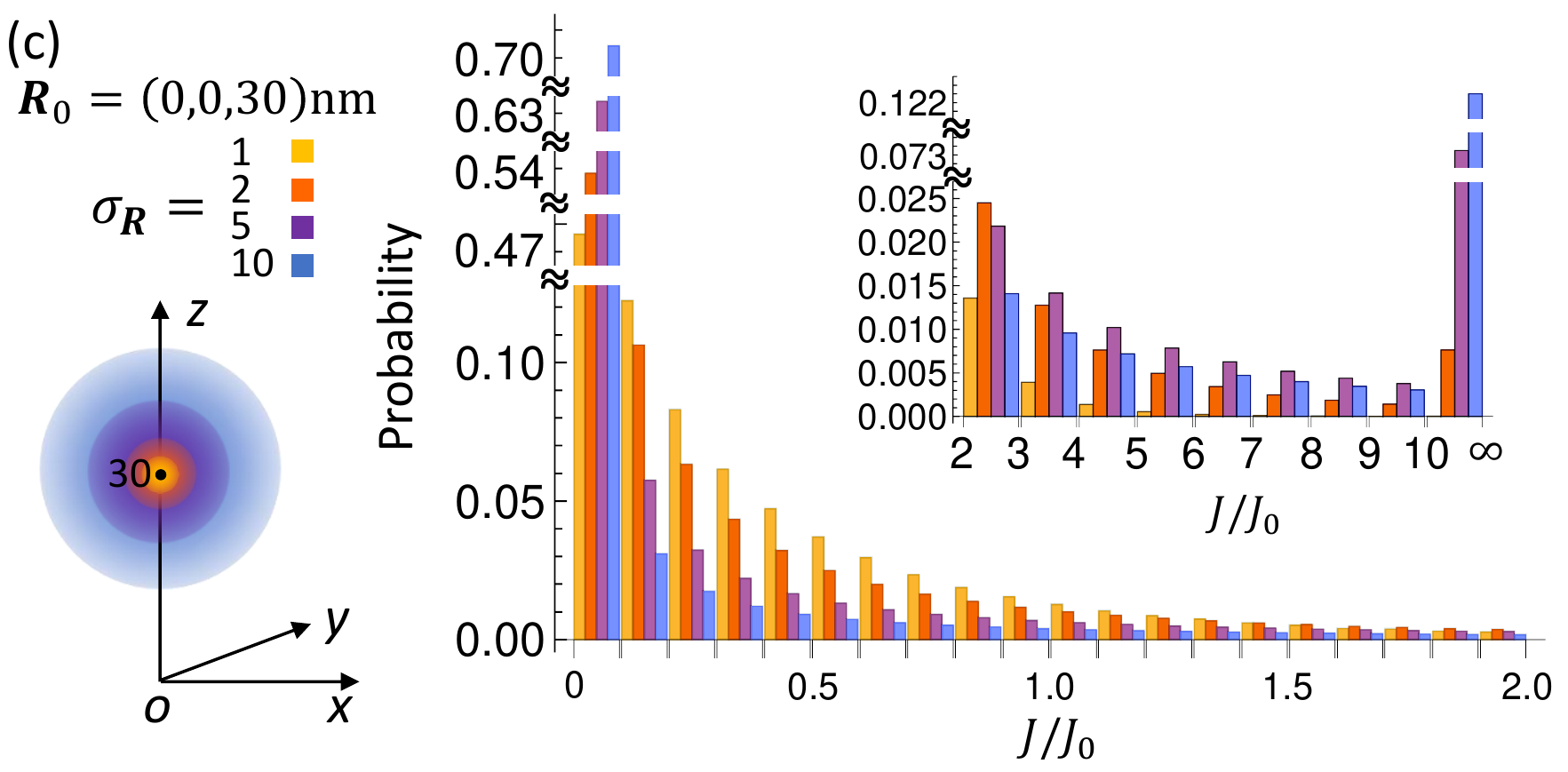}}
  \caption{ Probability distribution of $J(\mathbf{R})$ for $\mathbf{R}_0\|[001]$ and $R_0=10$ (a), 20 (b), or 30 nm (c), due to  donor placement uncertainty with the standard deviation $\sigma_R$. The uncertainty is drawn schematically on the left of each figure. The main probability histogram on the right is for $J\in [0,2J_0]$ with $\Delta J=0.1 J_0$, and the inset is for $J\in [2J_0, 10 J_0]$ with $\Delta J= J_0$ and for $J>10 J_0$.
  }
  \label{fig:fig_2}
\end{figure}

The straggling donor separation is modeled by the 3D Gaussian distribution, $(\sqrt{2\pi}\sigma_R)^{-3} \exp(-|\mathbf{R}-\mathbf{R}_0|^2/2\sigma_R^2)$, with the standard deviation $\sigma_R$ defining the straggle size. It approximately captures the similar lateral and longitudinal straggles in a large class of ion implantations \cite{Jamieson_APL05}, while a varied anisotropy would not change our results qualitatively. Instead of just calculating the chance of yielding $J(\mathbf{R})$ above a certain threshold, we analyze in full detail the probability over a relevant array of $J$ spectrum. Given this goal,  we examine representative cases for different $R_0$'s along all three  high-symmetry directions, over mutual straggles $\sigma_R$ ($=\sqrt{\sigma^2_1+\sigma^2_2}$, combining those of donors 1 and 2) from the smallest 1 nm to a relatively large (but still at the frontier of single-ion implantation technology) 10 nm.

Figure~\ref{fig:fig_2} shows the calculated exchange coupling probability distribution for $\mathbf{R}_0\| [001]$. The probability distribution here possesses a 8-fold axial symmetry, and as a result at most 6 wedges are explicitly integrated over. Still, to compute the probability of $J$ in each interval $\Delta J$, we use a high density grid ($\Delta R=0.0625$ nm) for the numerical integration to sort and filter the highly oscillatory function $J(\mathbf{R})$ into numerous unconnected regions. We study the detailed histogram for $J$ between $0.1 J_0$  and $2J_0$ with $\Delta J= 0.1 J_0$. This range is chosen just to be on the same order of magnitude of $J_0$, while $\Delta J$ is such that for each interval around $J_0$ the probability is non-negligible ($>1\%$). We note that these $\Delta J$ values are apparently far from quantum error correction thresholds (QECTs). To be complete, we also quantify the probabilities for $J>2J_0$ in the insets. We set $R_0=10, 20$ and 30 nm in Fig.~\ref{fig:fig_2} (a), (b) and (c) respectively and vary $\sigma_R=1, 2, 5$ and 10 nm (only the first three for $R_0=10$ nm; a less elaborate result corresponding to Fig.~\ref{fig:fig_2}(a) earlier appeared in Ref.~\onlinecite{Koiller_IEEE05}). Different $\sigma_R$'s are compared side-by-side in each $\Delta J$ interval, and clear trends can be read off varying $J$, $\sigma_R$ and $R_0$. With decreasing $J$, the probability ($P$) increases in a superlinear manner generally, even for the smallest straggle. As we will see, this is one of the most important consequences of the oscillatory-decay exchange coupling in Si, and it severely constrains the acceptable $\sigma_R$ values for building quantum computers. Compared among different $\sigma_R$'s, more weight near $J_0$ results for the smaller $\sigma_R$, and for larger $\sigma_R$ it gradually approaches the two extrema of $J$ values. This can be understood solely based on the exponential-varying envelope of $J(\mathbf{R})$,  as the bigger $\sigma_R$ sphere has more access to larger deviations from $J_0$ while at the same time the near-$J_0$ region occupies a \textit{relatively} smaller portion. Finally, different $R_0$'s bring comparatively the least change  to the $P_{\sigma\!_R}(J/J_0)$ profiles. Again this can be explained by the exponential variation, where within each $\sigma_R$ sphere ($\sigma_R$ is always chosen smaller than $R_0$) $J_{\sigma\!_R}(\mathbf{R})$ has very similar \textit{relative} variation around different $R_0$.

\begin{figure}[!htbp]
\centering
  \subfloat
  {\includegraphics[width=8.5cm]{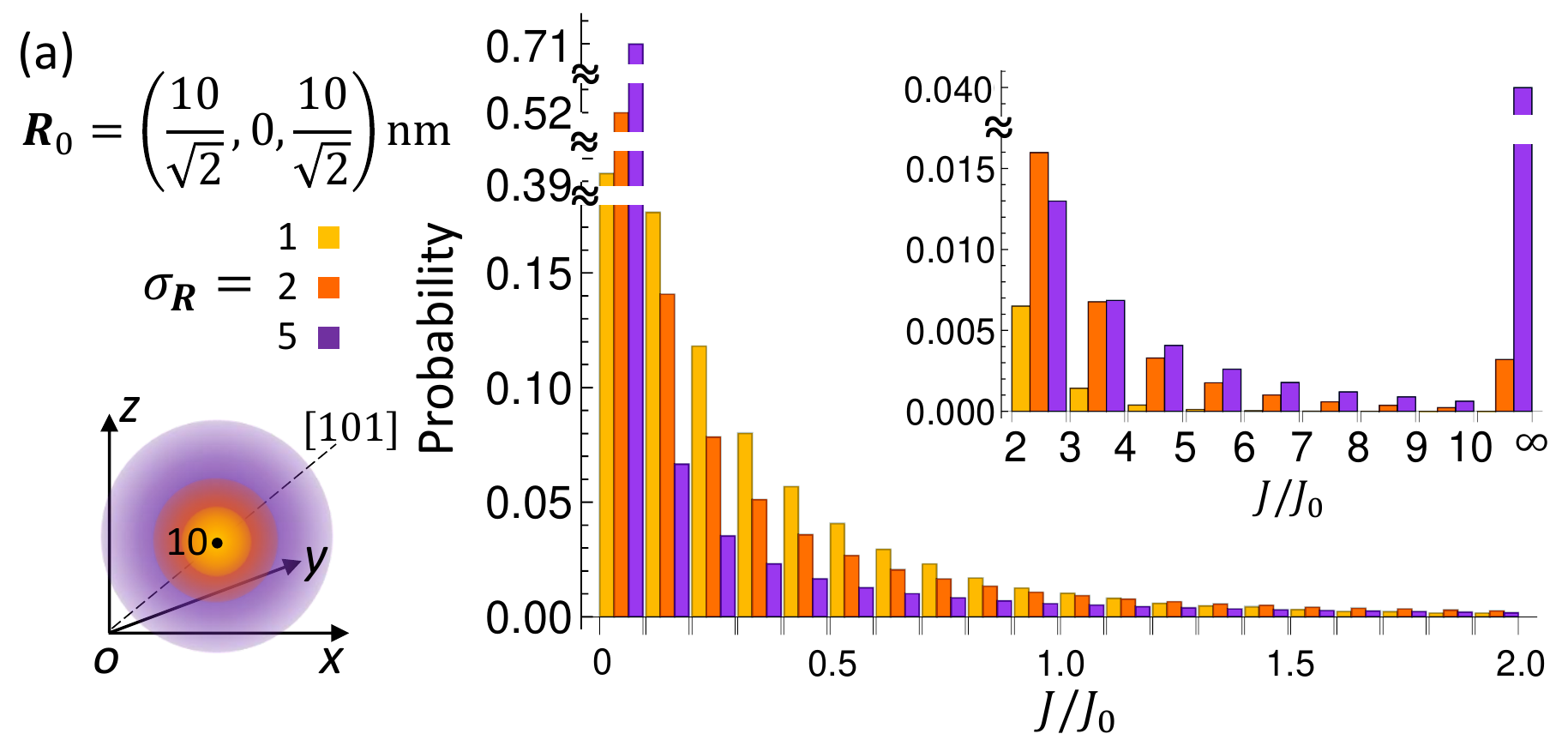}}
  \\
  \subfloat
  {\includegraphics[width=8.5cm]{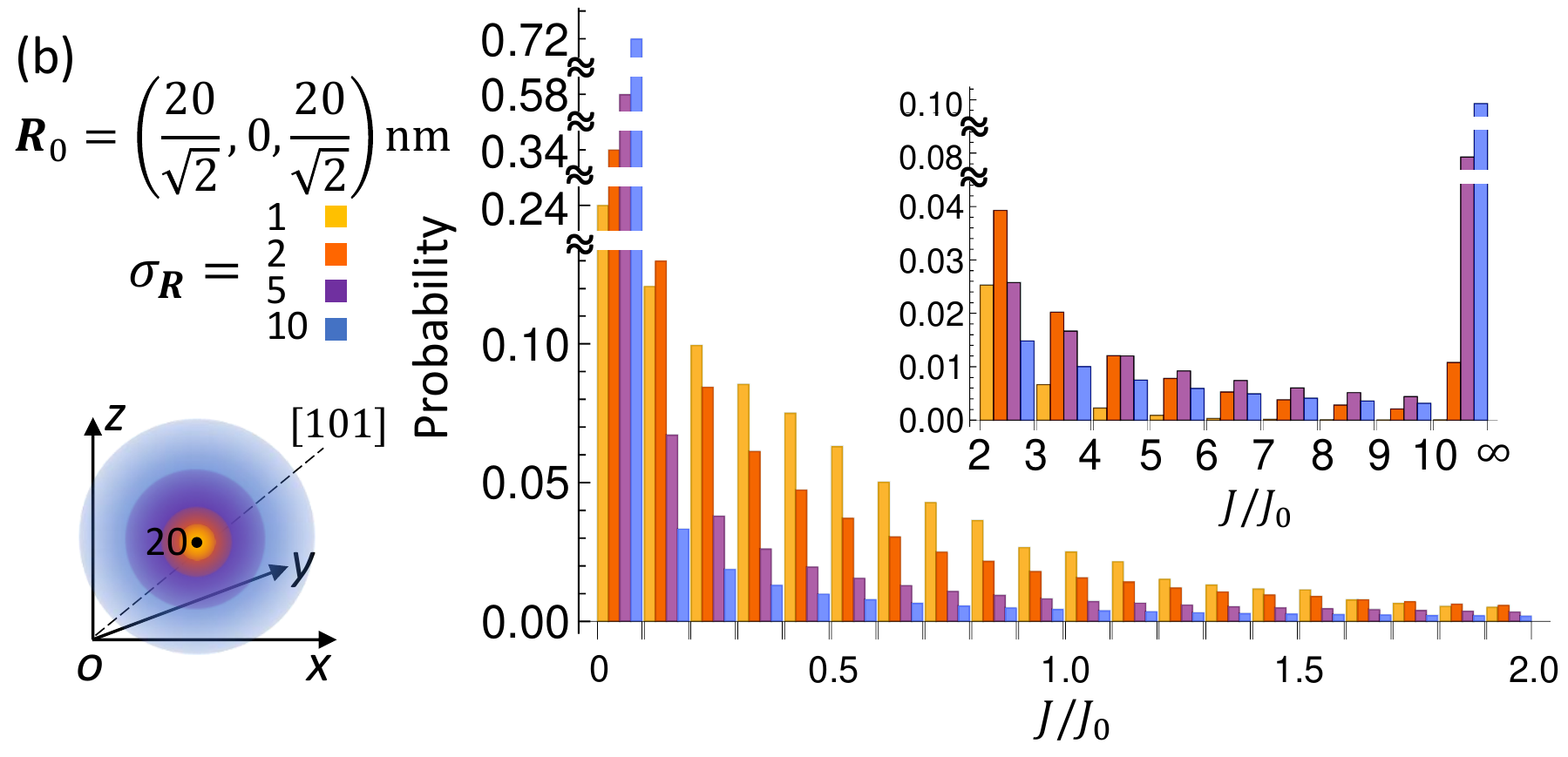}}
  \\
   \subfloat
  {\includegraphics[width=8.5cm]{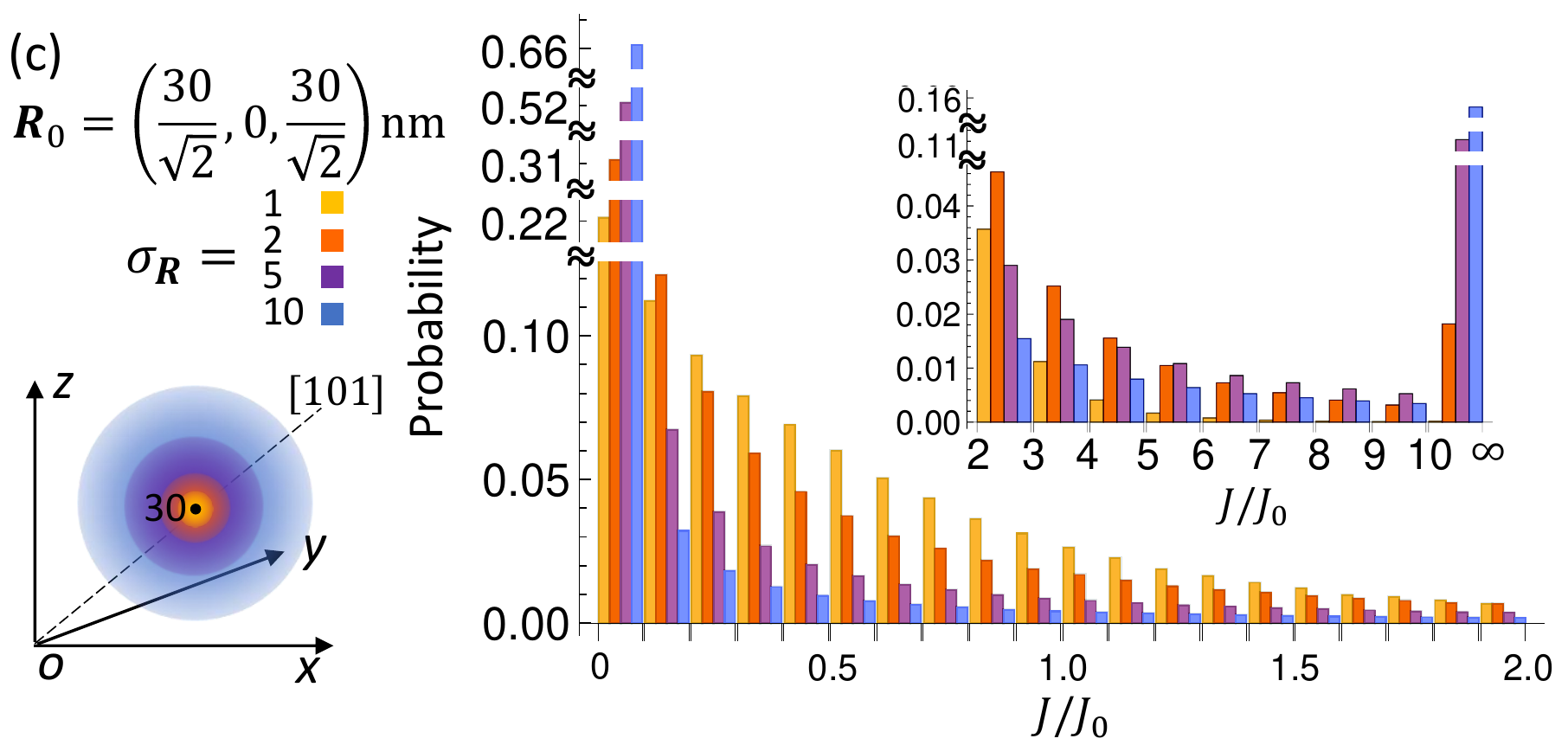}}
  \caption{Counterpart of Fig.~\ref{fig:fig_2} for $\mathbf{R}_0\|[101]$.}
  \label{fig:fig_3}
\end{figure}

\begin{figure}[!htbp]
\centering
  \subfloat
  {\includegraphics[width=8.5cm]{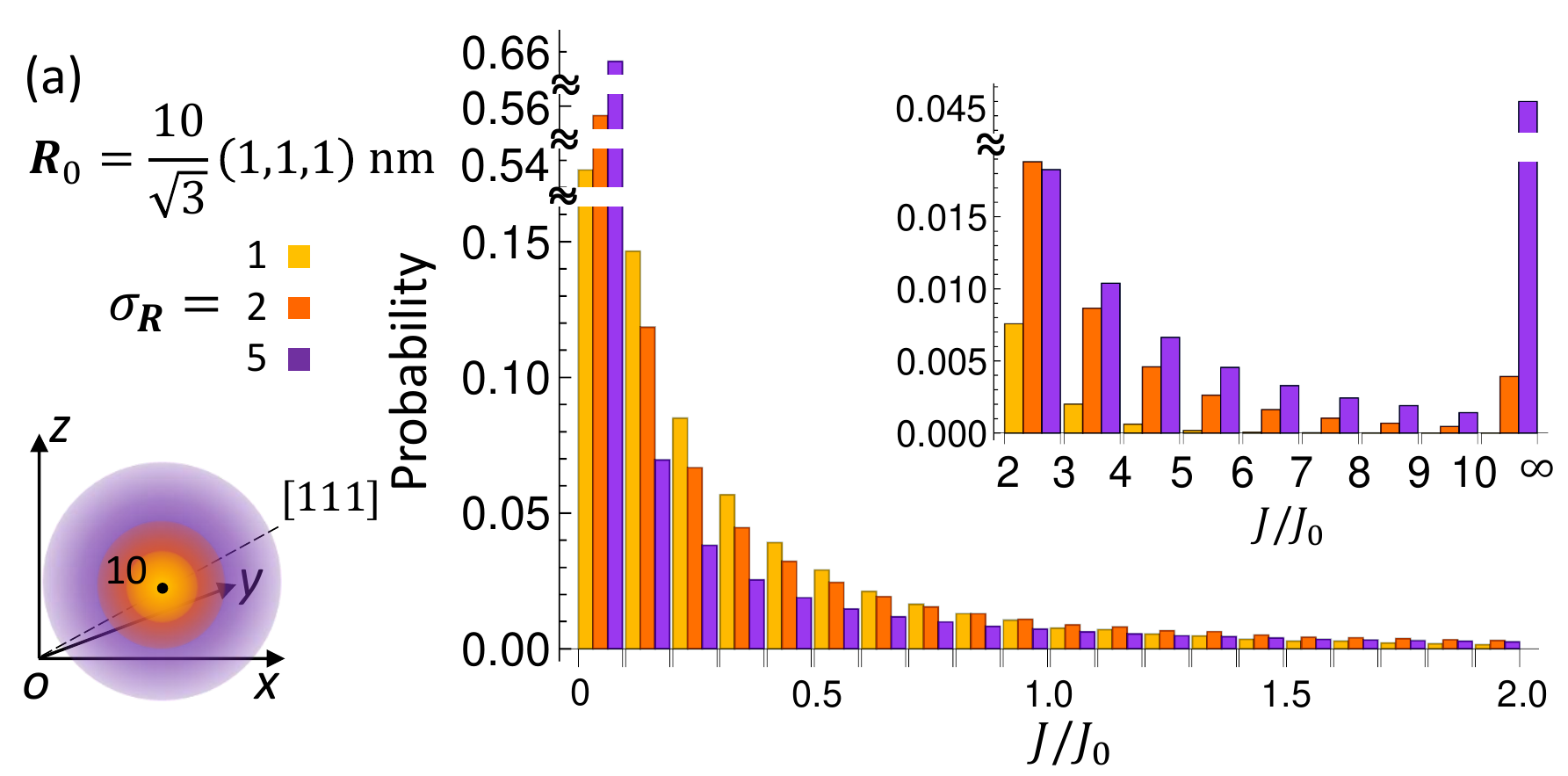}}
  \\
  \subfloat
  {\includegraphics[width=8.5cm]{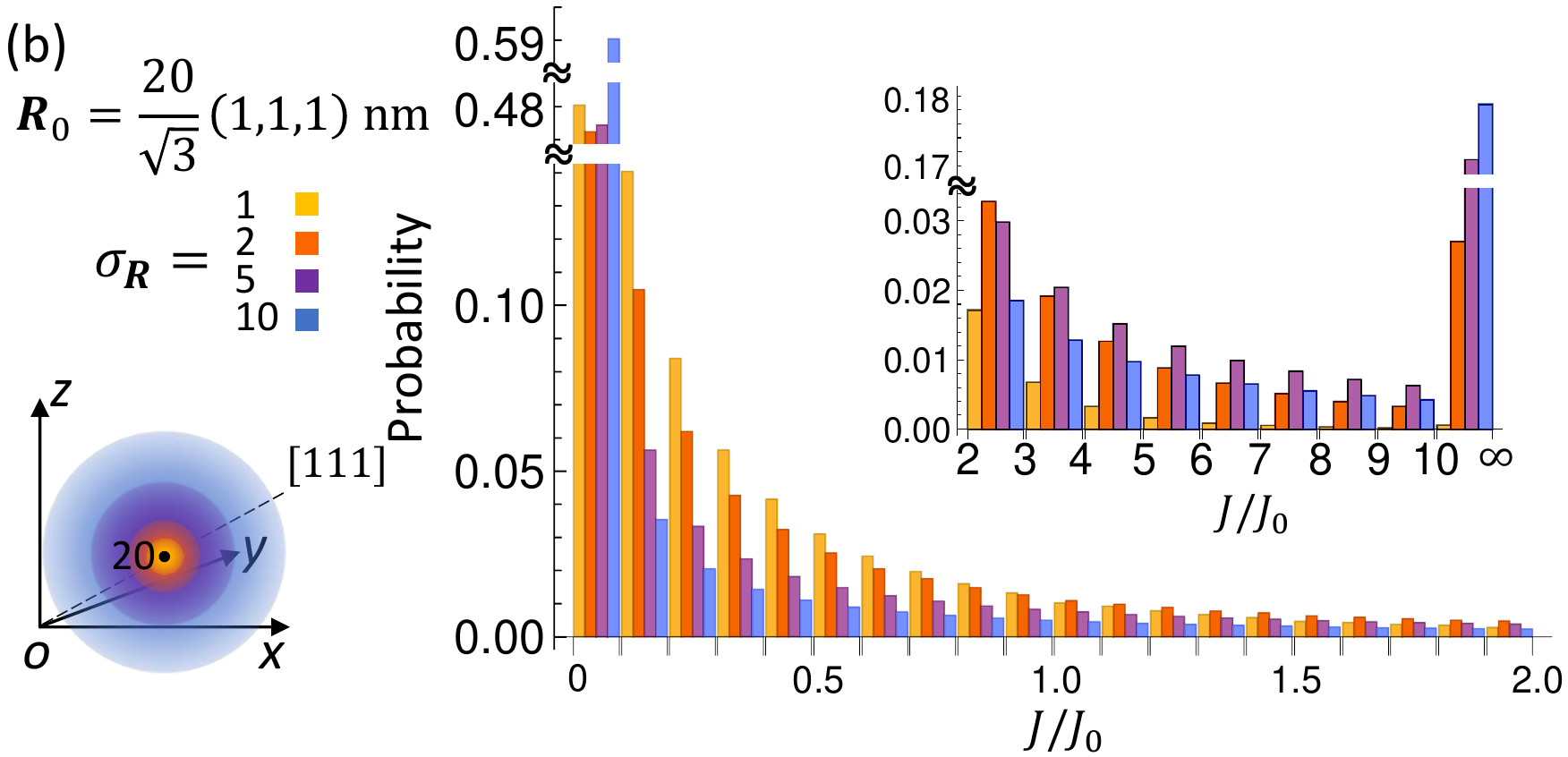}}
  \\
   \subfloat
  {\includegraphics[width=8.5cm]{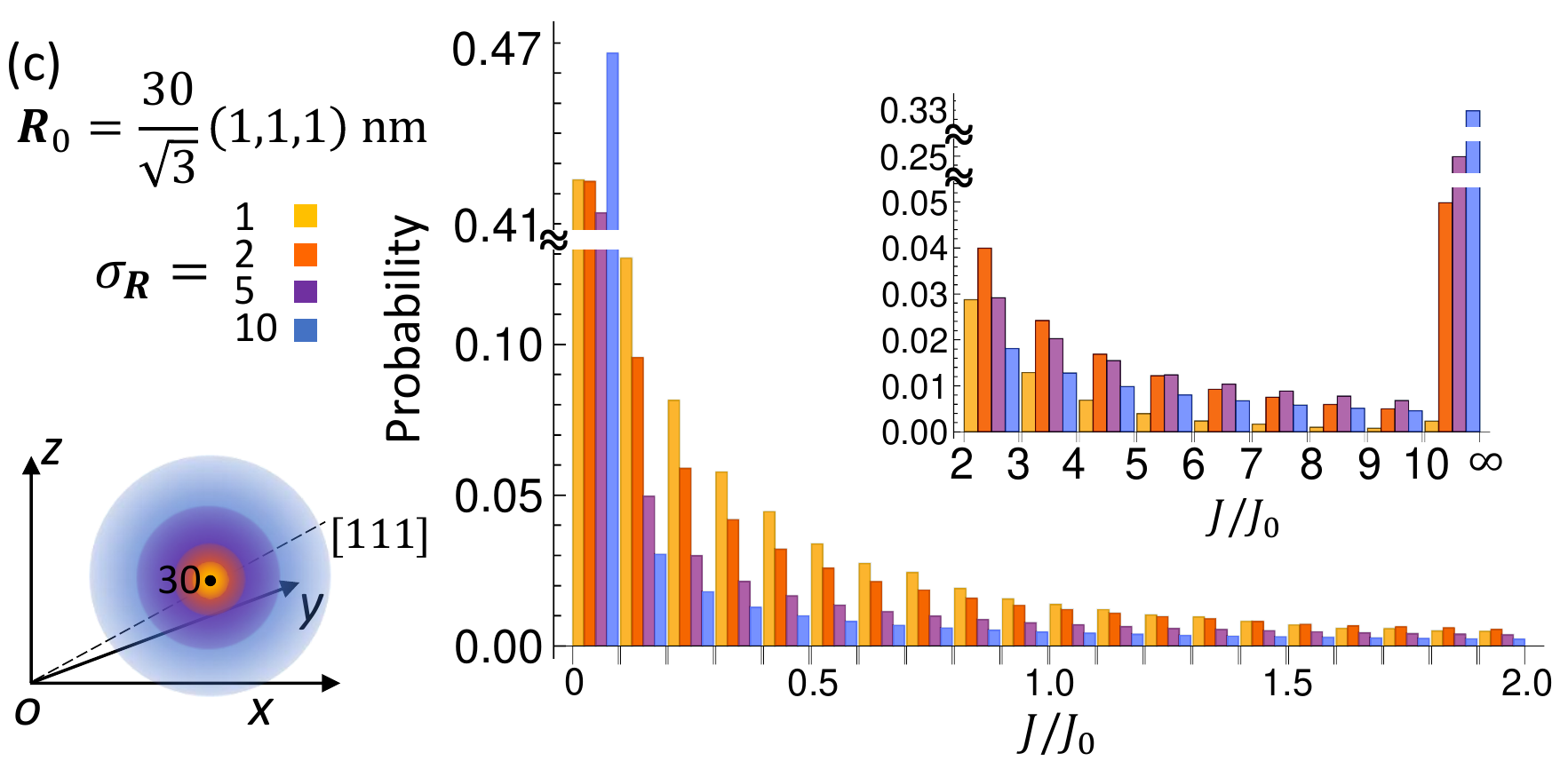}}
  \caption{Counterpart of Fig.~\ref{fig:fig_2} for $\mathbf{R}_0\|[111]$.}\label{fig:fig_4}
\end{figure}

The corresponding results for $\mathbf{R}_0\| [101]$ and [111] are presented in Figs.~\ref{fig:fig_3} and \ref{fig:fig_4}. Due to different symmetry of $J(\mathbf{R})$ around $\mathbf{R}_0$, 12 and 8 wedges are explicitly integrated over in these two respective cases. The first two trends described for the [001] case continue to hold. The dependence on $\mathbf{R}_0$ becomes pronounced as compared to the [001] case. The difference among [001], [101] and [111] (note each one represents a star of equivalent directions) arises from the anisotropy of the Si crystal field.  In general the resulting $J(\mathbf{R}\|[001])> J(\mathbf{R}\|[101]) >J(\mathbf{R}\|[111])$ for the same $|\mathbf{R}|$. As a result, for instance, going away from $\mathbf{R}_0\| [111]$, $J(\mathbf{R})$ manifests a transversal increase in additional to the longitudinal (radial) exponential change (on top of the oscillation). At $|\mathbf{R}|=30$ nm, $J(\mathbf{R})$ varies by more than 100 times. A direct consequence of this transverse anisotropy is that $P(J)$ decreases at the lower $J$ end ($[0, 0.1 J_0]$) and increases at the upper end from [001] to [101] and further to the [111] direction, for a fixed $\sigma_R$. Considering the change in both radial and angular directions in the [101] and especially [111] cases, a larger $R_0$ shifts more weight from the lower $J/J_0$ to the higher $J/J_0$ segments. This trend is less significant in the [001] case where it is perceivable only for $\sigma_R=5$ or 10 nm in the $J\in[10 J_0, \infty]$ segment.

\begin{figure}[!htbp]
\centering
\includegraphics[width=8.5cm]{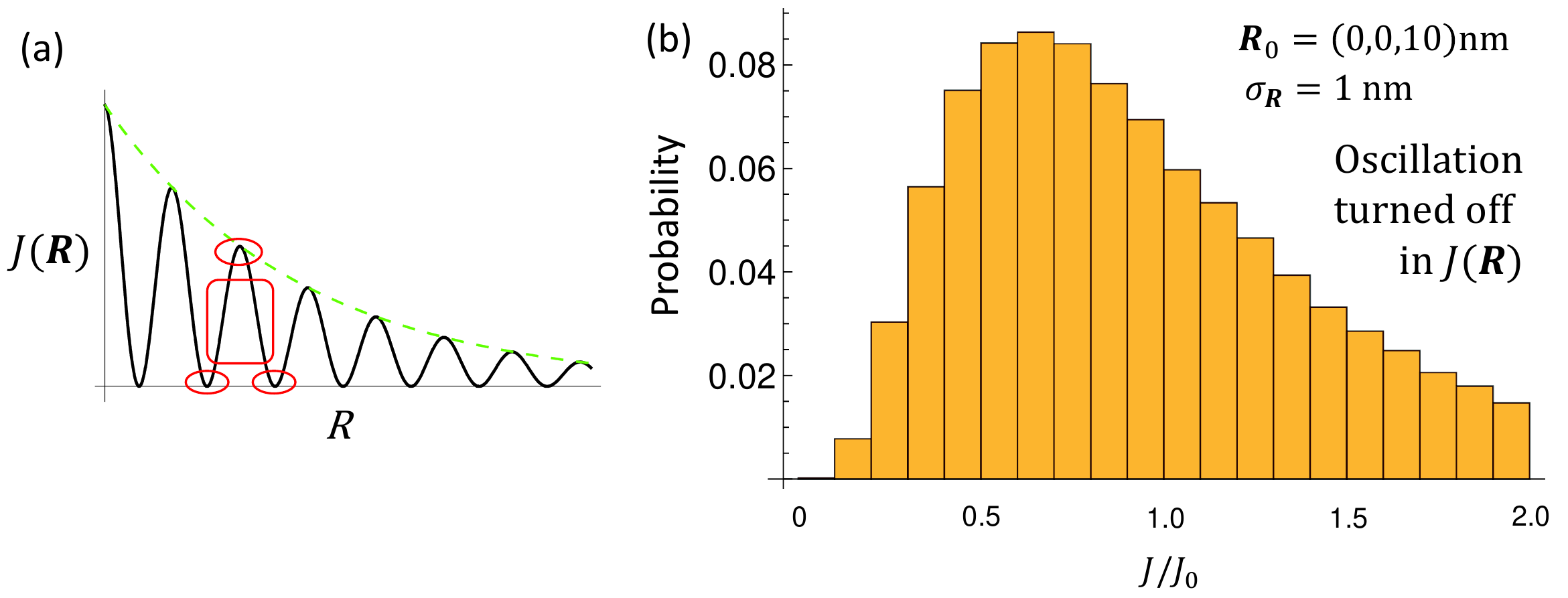}
\caption{ (a) A general schematic oscillatory decay $J(\mathbf{R})$, with its exponential envelope marked by the dashed line. The peak region, lower $J$ region, and the valley region in an arbitrary oscillation period are highlighted. The latter two regions gain finite probability, whereas they are absent for an exponential-only $J(\mathbf{R})$.  (b) The probability histogram  for  $\mathbf{R}_0=(0,0,10)$ nm, $\sigma_R=1$ nm between 0 and $2 J_0$, with the oscillation factors switched off in $J(\mathbf{R})$.
}\label{fig:fig_5}
\end{figure}

To appreciate the implications of the calculated straggle-induced  exchange coupling fluctuations in practice, we note that the generic exponential modulation of $J(\mathbf{R})$ with increasing separation and the specific exchange oscillation due to the Si conduction band valley degeneracy are the two fundamental mechanisms at play. Each one is characterized by a length scale, $r_B\sim 2$ nm for the former and $a\sim 0.5$ nm for the latter. The dominant behavior for a given straggle is determined by the straggling range with respect to these length scales. When it is  larger than $r_B$, such as $\sigma_R=5$ or 10 nm, the probability histogram on $0.1 J_0<J<10J_0$ is only slightly modified by the presence of oscillation, and has a higher weight towards lower $J$ intervals where the smaller $dJ/dR$ yields a larger $\mathbf{R}$ volume. This is clearly undesirable for achieving a narrow $J$ distribution. The oscillation in $J(\mathbf{R})$ becomes significantly relevant when the straggling range is between $a$ and $r_B$. It leads to a departure from the histogram of the usual exponential decay, and no matter how small the straggle is (still larger than lattice spacing), there is always a considerable probability at the low $J$ region ($J\ll J_0$). A schematic exponential-modulated oscillation is shown in Fig.~\ref{fig:fig_5} (a).  Especially for the valley region in each oscillation period, $dJ/dR\sim 0$, and thus the $\Delta J$ interval near $J=0$ occupies the largest volume and results in the tallest columns in Figs.~\ref{fig:fig_2}-\ref{fig:fig_4}. To drive this point home, Fig.~\ref{fig:fig_5}(b) shows the histogram result for $\mathbf{R}_0=(0,0,10)$ nm and a small $\sigma_R=1$ nm, after switching off the oscillatory factors in $J(\mathbf{R})$. In contrast to Fig.~\ref{fig:fig_2}(a), the exchange coupling peaks around $J_0$ with a width $\sim\sigma_R$. This lack of probability localization is another way of showing the detrimental and sensitive effect of oscillation on exchange coupling. Previous studies \cite{Wellard_PRB05, Pica_PRB14} indicate weakened oscillation of $J(\mathbf{R})$ by more refined theoretical models. We note that unless the uncertainty in donor separation has been made reliably below $r_B$,  the inevitable exponential variation is the practical limitation. Current ion implantation techniques provide straggle ranges much larger than the Si Bohr radius.

Even without exchange oscillation, however, the relative localization of $J$ and the `slow' exponential variation for $\sigma_R<r_B$ are still too problematic for any available QECT ($\leq 10^{-3}$ with reasonable overheads). This should be contrasted with the existing Si C-MOS microelectronics  technology where fairly large fluctuations can be tolerated around the target values even for devices which are only a few tens of nm in size. Therefore, for quantum computation we have the demanding (third) factor that the QECT tolerates only extremely small $J$ error. The QECT constraint makes almost any uncertainty in donor placement destructive whether or not exchange oscillation is present.

In conclusion, through concrete calculations and systematic analysis of two-qubit exchange coupling, we have emphasized the importance of precise donor placement (well beyond the current capability of commercial ion implantation techniques) in Si for the realization of scalable two-qubit gates in donor based quantum computing platforms. Without the ability to control $J$ precisely, qubit circuits beyond single qubit operations are likely to have unacceptable two-qubit gate errors beyond the quantum error correction thresholds. Our results for  $J$ error statistics provide important and clear lookup tables for experimentally relevant situations. The obvious solution is to develop fabrication techniques for reducing straggle below the lattice constant ($\sim$ 0.5 nm), but this is a demanding task using the available ion implantation techniques.  One possibility might be to move entirely to the so-called "bottom up" atomistic fabrication techniques (such as STM), but these are inherently new technologies whose ability for fabricating hundreds of qubits has not yet been demonstrated.  We mention that the problem discussed in our work does not in any way affect the feasibility of experimentally studying individual two-qubit gates in the laboratory where one can simply work with a device with two implanted donors as long as there is a measurable exchange coupling between the donor electrons.  Our analysis points to the difficulties in scaling up to many qubits where our calculated exchange probability distribution indicates that most scaled up devices will simply fail unless the straggle can be reduced by one or two orders of magnitude below that available currently.
Exchange coupling is determined by  donor separation as well as external electrical gating, and we have focused on the former. The placement error may be partially compensated  by gating (not eliminated, as the interference between multivalley wavefunctions is retained by the electrical field) which requires complex individual qubit calibration/control and array coordination. Moreover, the same mechanism also makes exchange coupling suffer from electrical noise. It is unclear that this gating strategy can be achieved in scaled up devices with many qubits, but of course, this is easy to do in a single device with just two donors.

This work is supported by LPS-MPO-CMTC.

\end{document}